\documentclass{elsart}

\usepackage{natbib}

\usepackage{epsfig}

\usepackage{amssymb}
\usepackage{amsbsy}

\begin{document}

\begin{frontmatter}



\title{Radiation from Relativistic Shocks in Turbulent Magnetic Fields}    %\thanksref{footnote1}}
%\thanks[footnote1]{This template can be used for all publications in Advances in Space Research.}

\author[label1]{K.-I. Nishikawa},
\author[label2]{J. Niemiec},
\author[label3]{M. Medvedev},
\author[label4]{B. Zhang},
\author[label5]{P. Hardee},
\author[label6]{\AA. Nordlund},
\author[label6]{J. Frederiksen},
\author[label1]{Y. Mizuno},
\author[label7]{H. Sol},
\author[label8]{M. Pohl},
\author[label9]{D. H. Hartmann},
\author[label10]{M. Oka}, and  
\author[label11]{G. J. Fishman}

\address[label1]{National Space Science and Technology Center,
  Huntsville, AL 35805, USA}
\address[label2]{Institute of Nuclear Physics PAN, ul. Radzikowskiego 152, 31-342 Krak\'{o}w, Poland}
\address[label3]{Department of Physics and Astronomy, University of Kansas, KS
66045, USA}
\address[label4]{Department of Physics, University of Nevada, Las
Vegas, NV 89154, USA} 
\address[label5]{Department of Physics and Astronomy,
  The University of Alabama, Tuscaloosa, AL 35487, USA} 
\address[label6]{Niels Bohr Institute, University of Copenhagen, 
Juliane Maries Vej 30, 2100 Copenhagen \O, Denmark}
\address[label7]{LUTH, Observatore de Paris-Meudon, 5 place Jules Jansen, 92195 Meudon Cedex, France}
\address[label8]{Department of Physics and Astronomy, Iowa State University, Ames, IA 50011, USA}
\address[label9]{Department of Physics and Astronomy, Clemson University, Clemson, SC 29634, USA}
\address[label10]{1Space Sciences Laboratory, University of California, Berkeley, California 94720, USA}
\address[label11]{NASA/MSFC,
  Huntsville, AL 35805, USA} 

%\author{First Author\corauthref{cor}\thanksref{footnote2}}
%\address{Address of the first author}
%\corauth[cor]{Corresponding author}
%\thanks[footnote2]{Additional information regarding the corresponding author}
%\ead{corresponding-author@email.address}
%url can be given like this
%\ead[url]{http://authors.elsevier.com/locate/latex}

\begin{abstract}
% Text of abstract

Using our new 3-D relativistic particle-in-cell (PIC) code parallelized with MPI, we investigated long-term particle
acceleration associated with a relativistic electron-positron jet propagating in an unmagnetized ambient electron-positron plasma. The simulations were performed using a much longer simulation system
than our previous simulations in order to investigate the full nonlinear stage of the Weibel instability and its particle acceleration mechanism. Cold jet electrons are thermalized and ambient electrons are accelerated in the resulting shocks. Acceleration of ambient electrons leads to a maximum ambient electron density three
times larger than the original value as predicted by hydrodynamic shock compression. In the jet (reverse) shock behind the bow (forward) shock the strongest electromagnetic fields are generated. These fields may
lead to time dependent afterglow emission. In order to calculate radiation from first principles that goes beyond the standard synchrotron model used in astrophysical objects we have used PIC simulations. Initially we calculated radiation from electrons propagating in a uniform parallel magnetic field to verify the 
technique. We then used the technique to calculate emission from electrons in a small simulation system. From these simulations we obtained spectra which are consistent with those generated from electrons propagating
in turbulent magnetic fields with red noise. This turbulent magnetic field is similar to the magnetic field generated at an early nonlinear stage of the Weibel instability. A fully developed shock within a larger simulation system may generate a jitter/synchrotron spectrum.  

\end{abstract}


\end{frontmatter}

\parindent=0.5 cm
\parskip 10pt

% main text
\vspace{-0.9cm}
\section{Introduction}
\vspace{-0.5cm}

Particle-in-cell (PIC) simulations can shed light on the physical mechanism of particle acceleration that occurs in the complicated dynamics within relativistic shocks.  Recent PIC simulations of relativistic electron-ion and electron-positron jets injected into an ambient plasma show that acceleration occurs within the downstream jet
\citep{nishi03,nishi05,hede05,nishi06,ram07,chang08,anat08a,anat08b,sironi09m}. In general, these  simulations have confirmed that the Weibel instability, which generates current filaments and associated
magnetic fields mediates the relativistic collisionless shock \citep{medv99}, and accelerates electrons 
\citep{hede05,nishi06,ram07,chang08,anat08a,anat08b,sironi09m}. Therefore, the investigation of radiation resulting from accelerated particles (mainly electrons and positrons) in turbulent magnetic fields is essential to 
understanding the radiation and observable spectral properties. In this article we present a numerical method to obtain spectra from particles self-consistently traced in our PIC simulations. 

\vspace{-0.7cm}
\section{Relativistic PIC Simulations}
\vspace{-0.5cm}

\subsection {An Unmagnetized Pair Jet Injected into an Unmagnetized Pair Plasma}
\vspace{-0.5cm}

We have performed a simulation using a system with ($L_{\rm x}, L_{\rm y}, L_{\rm z}) = (4005\Delta,$ $131\Delta, 131\Delta)$ ($\Delta = 1$: grid size) and a total of $\sim 1$ billion particles (12 particles$/$cell$/$species for the ambient plasma) in the active grid zones \citep{nishi09}.  In the simulation the electron skin depth, $\lambda_{\rm ce} = c/\omega_{\rm pe} = 10.0\Delta$, where $\omega_{\rm pe} = (4\pi e^{2}n_{\rm
e}/m_{\rm e})^{1/2}$ is the electron plasma frequency and the electron Debye length $\lambda_{\rm e}$ is half of the grid size. Here the computational domain is six times longer than in our previous simulations \citep{nishi06,ram07}.  The electron number density of the jet is $0.676n_{\rm e}$, where $n_{\rm e}$ is the ambient electron density and $\gamma = 15$.   This parameter regime may be relevant to gamma-ray burst
afterglows and AGN jets. The electron/positron thermal velocity of the jet is $v^{\rm e}_{\rm j,th} = 0.014c$, where $c = 1$ is the speed of light. 

%Jets are injected in a plane across the computational grid at
%$x = 25\Delta$ in the positive $x$ direction in order to eliminate
%effects associated with the boundary conditions at $x = x_{\rm
%\min}$. Radiating boundary conditions were used on the planes at {\it
%$x = x_{\min}~{\&}~x_{\max}$}. Periodic boundary conditions were used
%on all transverse boundaries.  The ambient and jet electron-positron
%plasma has mass ratio $m_{\rm e^-}/m_{\rm e^+} = 1$.
%The electron/positron thermal velocity in the ambient plasma is 
%$v^{\rm e}_{\rm a,th} = 0.05c$. 

Figure 1 shows the averaged (in the $y-z$ plane) electron density and electromagnetic field energy along the jet at  $3750\omega_{\rm pe}^{-1}$.  The resulting profiles of jet (red), ambient (blue), and total (black) electron 
density are shown in Fig.\ 1a. 
\begin{figure}[h]
\begin{center}
\includegraphics[width=200pt, height=200pt]{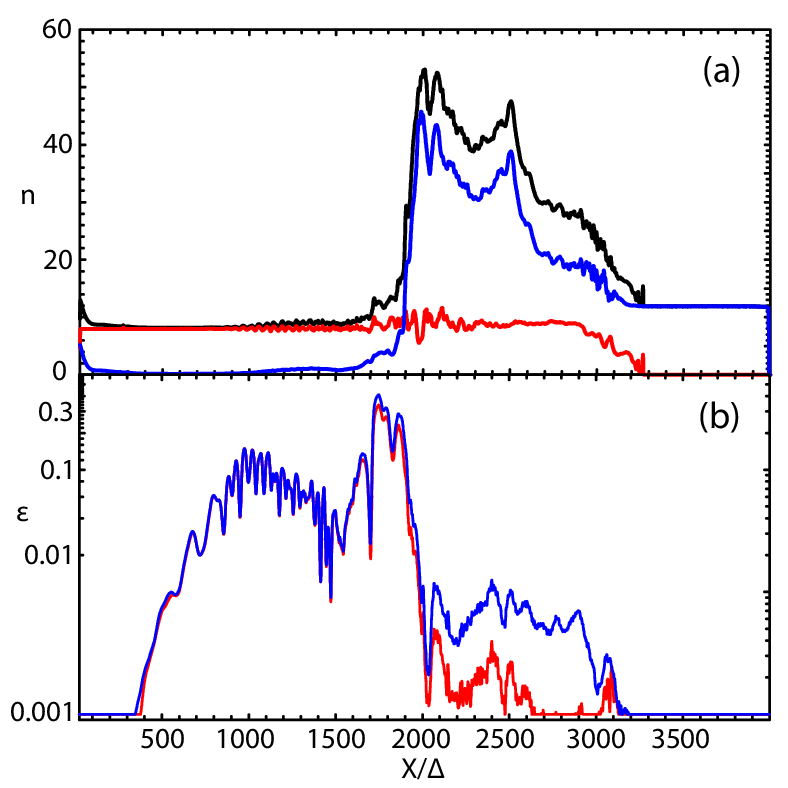}
\end{center}
%\vspace*{-12.0cm}
\caption{The averaged values of electron density (a) and field energy (b) along the $x$ at $t = 3750\omega_{\rm pe}^{-1}$. Fig.\ 1a shows jet electrons (red), ambient electrons (blue), and the total electron density (black). Fig.\ 1b shows electric field energy (red) and magnetic field energy (blue) divided by the total kinetic energy.} 
\end{figure}
The ambient electrons are accelerated by the jet electrons and pile up towards the front part of jet.  At earlier times the ambient plasma density increases linearly behind the jet front. At the later times the ambient plasma
shows a rapid increase to a plateau behind the jet front, with additional increase to a higher plateau farther behind the jet front. The jet density remains approximately constant except near the jet front.

The Weibel instability remains excited by the continuously injected jet particles and the electromagnetic fields are maintained at a high level, about four times that seen in a previous, much shorter grid simulation system (with $L_{\rm x} = 640\Delta$). At earlier simulation times  a large electromagnetic structure is generated and accelerates the ambient plasma. As shown in Fig.\ 1b, at this later simulation time the strong magnetic field 
extends up to $x/\Delta = 2,000$. These strong fields become small  in the shocked ambient region beyond $x/\Delta = 2000$ \citep{nishi06,ram07}.

The acceleration of ambient electrons becomes visible when jet electrons pass about $x/\Delta = 500$. The maximum density of accelerated ambient electrons is attained at $t = 1750\omega_{\rm pe}^{-1}$. The maximum density gradually reaches a plateau as seen in Fig.\ 1a. The maximum electromagnetic field energy is located at $x/\Delta = 1,700$ as shown in Fig.\ 1b. 

\vspace{-0.5cm}
\subsection{A Numerical Method for Calculating Emission}
\vspace{-0.5cm} 

Let a particle be at position ${\bf{r}_{0}}(t)$ at time $t$  \citep{nishi08,hedeT05,hedeN05}. At the same time, 
we observe the electric field from the particle from position $\bf{r}$. However, because of the finite velocity of light, we observe the particle at an earlier position $\bf{r}_{0}(\rm{t}^{'})$ where it was at the retarded time $t^{'} = t - \delta t^{'} = t - \bf{R}(\rm{t}^{'})/c$. Here $\bf{R}(\rm{t}^{'}) = |\bf{r} - \bf{r}_{0}(\rm{t}^{'})|$ is the distance from the charge (at the retarded time $t^{'}$) to the observer.

After some calculation and simplifying assumptions the total energy $W$ radiated per unit solid angle per unit frequency from a charged particle moving with instantaneous velocity $\boldsymbol{\beta}$ under
acceleration $\boldsymbol{\dot{\beta}}$ can be expressed as \citep{rybi79,jack99}
\begin{eqnarray}
\frac{d^{2}W}{d\Omega d\omega} & =  &
 \frac{\mu_{0} cq^{2}} {16\pi^{3}} \left|\int^{\infty}_{-\infty}\frac{\bf{n}\times
[(\bf{n}-\boldsymbol{\beta})\times \boldsymbol{\dot{\beta}}]}{(1-\boldsymbol{\beta}\cdot
\bf{n})^{2}} e^{i\omega(t^{'} -\bf{n} \cdot \bf{r}_{0}({\rm t}^{'})/{\rm c})}
dt^{'}\right|^{2}  
\end{eqnarray}
Here, $\bf{n} \equiv \bf{R}(\rm{t}^{'})/ |\bf{R}(\rm{t}^{'})|$ is a unit vector that points from the particle's retarded position towards the observer. 

The observer's viewing angle is set by the choice of $\bf{n}$ ($n_{\rm x}^{2}+n_{\rm y}^{2}+n_{\rm z}^{2} = 1$). 
The choice of unit vector $\bf{n}$ along the direction of propagation of the jet (hereafter taken to be the $x$-axis) corresponds to head-on emission. For any other choice of $\bf{n}$ (e.g., $\theta_{\gamma} = 1/\gamma = (1 - \beta^{2})^{1/2} = (1 -  (\boldsymbol{v}/c)^{2})^{1/2}$), off-axis emission is seen by the observer.  It is noted that in this article that radiative losses are not included
 \citep[e.g.,][]{jaro09,rev10}.

In order to calculate radiation from relativistic jets propagating along the $x$-direction \citep{nishi08} we consider a test case which includes a parallel magnetic field ($B_{\rm x}$), and jet velocity of $v_{\rm j1,2} = 0.99c$. Two electrons are injected with different perpendicular velocities ($v_{\perp 1} = 0.1c, v_{\perp 2} 
= 0.12c$). A maximum Lorentz factor of $\gamma_{\max} =\{(1 - (v_{\rm j2}^{2} +v_{\perp 2}^{2})/c^{2}\}^{-1/2}  
= 13.48$  is calculated for the larger perpendicular velocity.

Figure 2 shows electron trajectories in the $x - y$ plane (red: $v_{\perp 2} = 0.12c$, blue: $v_{\perp 1} = 0.1c$) (a: left panel), the radiation (retarded) electric field  (b: middle panel), and spectra (c: right panel) for $B_{\rm x} = 3.70$.  The two electrons are propagating from left to right with gyration in the $y - z$ plane (not
shown). The gyroradius is about $0.44\Delta$ for the electron with the larger perpendicular velocity. 
The power spectrum is shown at seven viewing angles  with respect to the $x$-direction of 0$^{\circ}$ (red), 10$^{\circ}$ (orange), 20$^{\circ}$ (yellow), 30$^{\circ}$ (moss green), 45$^{\circ}$ (green), 70$^{\circ}$ (light blue), and 90$^{\circ}$ (blue). The higher frequencies become stronger at the $10^{\circ}$ viewing angle. The critical angle for off-axis radiation $\theta_{\gamma} = 180^{\circ}/(\pi \gamma_{\max})$ for this case is
4.25$^{\circ}$. As shown in this panel, the spectrum at a larger viewing angle ($> 20^{\circ}$) has smaller amplitude.
\begin{figure}[h]
\begin{center}
\includegraphics[width=110pt, height=110pt]{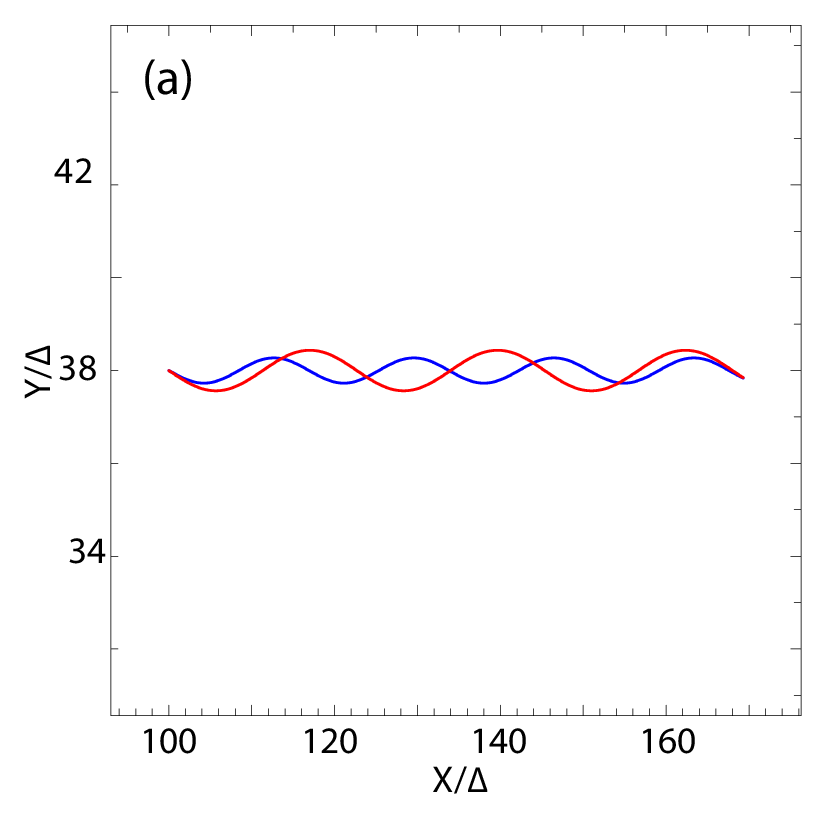}
\hspace*{-0.3cm}
\includegraphics[width=150pt, height=60pt]{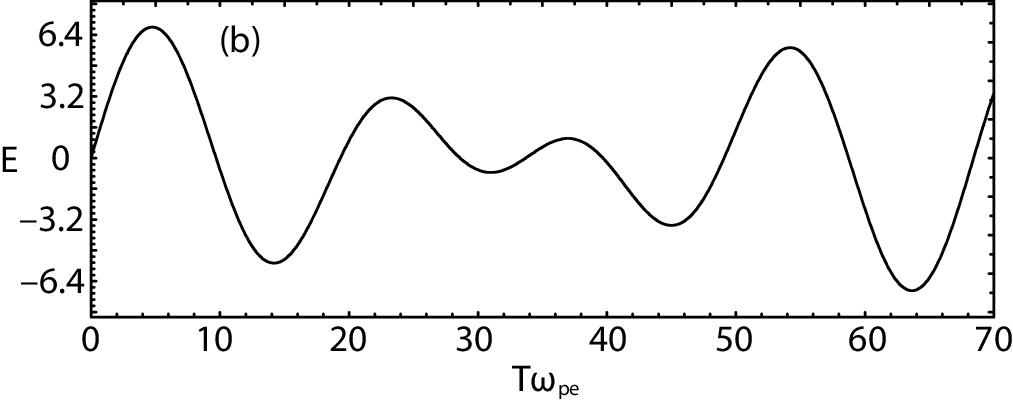}
\hspace*{-0.2cm}
\includegraphics[width=120pt, height=120pt]{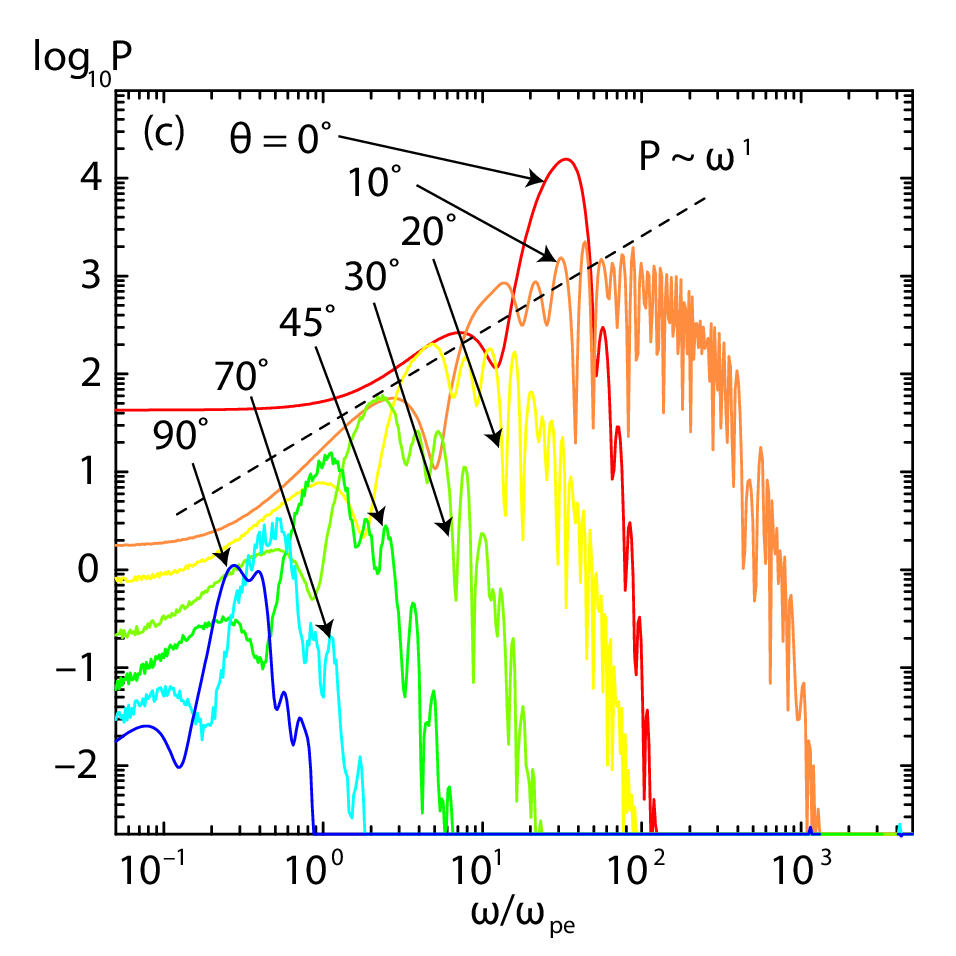}
\end{center}
\vspace*{-0.0cm}
\caption{\baselineskip 12pt A test case with a strong magnetic field ($B_{\rm x} = 3.7$) and two electrons
 with different perpendicular velocities  ($v_{\perp 1} = 0.1c, v_{\perp 2}  = 0.12c$). (a) The paths of the helically moving electrons along the $x-$direction in the homogenous magnetic field are shown in the $x-y$-plane. (b) The two electrons radiate a time dependent electric field. The retarded electric field from the moving electrons seen by an observer situated in the rest frame at great distance along the n-vector. (c) The observed power spectrum at different viewing angles from the two electrons. Frequency is in units of $\omega_{\rm pe}^{-1}$. }
\end{figure}
 Only two different electrons are used to calculate the radiation, therefore two cyclotron frequencies and their higher harmonics are visible in Fig.\ 2c, and the spectra are not smooth.  Since the jet plasma has a large velocity $x$-component in the simulation frame, the radiation from the particles (electrons and positrons) is strongly beamed along the $x$-axis (as in jitter radiation) \citep{eps73,medv00,medv06}. 

Equations 6.30a and 6.30b in \citet{rybi79} are an approximation suggesting that radiation at the viewing angle $\alpha = 0^{\circ}$ (eq. 6.33) disappears (see Fig. 6.5 in the textbook of  \citet{rybi79}). However, an exact expression shows that radiation at viewing angle $0^{\circ}$ does not vanish \citep{beke66,land80}  as a result of the relativistic distortion of the lobes of a Doppler-boosted dipole antenna pattern. This aspect is shown in Fig.\ 2c, and note that the amplitude at higher frequency at the viewing angle $10^{\circ}$ is stronger than at viewing angle $0^{\circ}$.

\vspace{-0.5cm} /Hoshino
\subsection{The Standard Synchrotron Radiation Model}
\vspace{-0.5cm}

A synchrotron shock model is widely adopted to describe the radiation  thought to be responsible for observed broad-band GRB afterglows \citep{zm04,p05a,p05b,zhangr07,nakar07}.
Associated with this model are three major assumptions that are adopted in almost all current GRB afterglow models. Firstly, electrons are assumed to be ``Fermi'' accelerated at the relativistic shocks and to have a
power-law distribution with a power-law index $p$ upon acceleration, i.e., $N(E_{\rm e})dE_{\rm e} \propto E^{-p} dE_{\rm e}$. This is consistent with recent PIC simulations of shock formation and particle acceleration \citep{anat08b} and also some Monte Carlo models \citep{Achterberg01,ellison02,lemoi03}, but see 
%Niemiec \& Ostrowski 2006, Niemiec et al. 2006
\citep{niemiec06,niemiecp06}. Secondly, a fraction $\epsilon_{\rm e}$ (generally taken to be $\leq 1$) of the electrons associated with ISM baryons are accelerated, and the total electron energy is a fraction $\epsilon_{\rm e}$ of the total internal energy in the shocked region. Thirdly, the strength of the magnetic fields in the shocked region is unknown, but its energy density ($B^{2}/8\pi$) is assumed to be a fraction $\epsilon_{B}$ of the internal energy. These assumed ``micro-physics'' parameters, $p, \epsilon_{\rm e}$ and $\epsilon_{\rm B}$, whose values are obtained from spectral fits \citep{paku01,yost03} reflect the lack of a detailed description of the microphysics \citep{wax06}. 

The typical observed emission frequency from an electron with (comoving) energy $\gamma_{\rm e}m_{\rm e}c^{2}$ in a frame with a bulk Lorentz factor $\Gamma$ is $\nu = \Gamma\gamma_{\rm e}^{2}(eB/2\pi m_{\rm e}c)$. Three critical frequencies are defined by three characteristic electron energies. These are $\nu_{\rm m}$
(the injection frequency), $\nu_{\rm c}$ (the cooling frequency), and $\nu_{\rm M}$ (the maximum synchrotron frequency). In our simulations of GRB afterglows, there is one additional relevant frequency, $\nu_{\rm a}$, due to synchrotron self-absorption at lower frequencies \citep{mrw98,spn98,zhangr07,nakar07}.

%(Meszaros, Rees, \& Wijers 1998; Sari, Piran, \& Narayan 1998; Nakar
% 2007; Zhang 2007).

The general agreement between blast wave dynamics and direct measurements of the fireball size argue for the validity of this model \citep{zhangr07,nakar07}. 
%(e.g. Zhang 2007; Nakar 2007).
The shock is most likely collisionless, i.e., mediated by plasma instabilities \citep{wax06}. 
%(Waxman 2006). 
The electromagnetic instabilities mediating the afterglow shock are expected to generate magnetic
fields. Afterglow radiation was therefore predicted to result from synchrotron emission of shock accelerated electrons \citep{mr97}. 
%(Meszaros \& Rees 1997). 
The observed spectrum of afterglow radiation is indeed remarkably consistent with synchrotron emission of electrons accelerated to a power-law distribution, providing support for the standard afterglow model based on synchrotron emission of shock accelerated electrons \citep{p99,p00,p05a,zm04,mes02,mes06,zhangr07,nakar07}.

%(Piran 1999, 2000, 2005a; Zhang \& Meszaros 2004; 
%Meszaros 2002, 2006; Zhang 2007; Nakar 2007).

In order to determine the luminosity and spectrum of synchrotron radiation, the strength of the magnetic field ($\epsilon_{\rm B}$) and  the index of the electron energy distribution  ($p$) must be determined. Due to the lack of a first principles theory of collisionless shocks, a purely phenomenological approach to the
model of afterglow radiation was ascribed without investigating in detail the processes responsible for particle acceleration and magnetic field generation \citep{wax06}. 
%(Waxman 2006). 
%Rather, one simply assumes
%that a fraction $\epsilon_{\rm B}$ of the post-shock thermal energy
%density is carried by the magnetic field, that a fraction
%$\epsilon_{\rm e}$ is carried by electrons, and that the energy
%distribution of the electrons is a power-law, $d \log n_{\rm e}/d
%\log \varepsilon = p$ (above some minimum energy $\varepsilon_0$
%which is determined by $\epsilon_{\rm e}$ and $p$), $\epsilon_{\rm
%B}$, $\epsilon_{\rm e}$ and $p$ are treated as free parameters, 
%determined by observations. 
It is important to clarify here that the constraints implied on these parameters by the observations are
independent of any assumptions regarding the nature of the afterglow shock and the processes responsible for particle acceleration or magnetic field  generation. Any model should satisfy these observational constraints.

The properties of synchrotron (or ``jitter") emission from relativistic shocks will be determined by the magnetic field strength and structure and the electron energy distribution behind the shock. The characteristics of jitter radiation may be important to understanding the complex time evolution and/or spectral structure in
gamma-ray bursts \citep{pre98}. 
%(Preece et al.\ 1998). 
For example, jitter radiation has been proposed as a means to explain GRB spectra below the peak frequency that are  harder than the ``line of death'' spectral index associated with synchrotron emission \citep{eps73,medv00,medv06},  i.e., the observed spectral power scales as $F_{\nu} \propto \nu^{2/3}$, whereas 
synchrotron spectra are $F_{\nu} \propto \nu^{1/3}$ or softer \citep{medv06}. Thus, it is essential to calculate radiation production by tracing electrons (positrons) in self-consistently generated small-scale electromagnetic fields.

\vspace*{-0.5cm}
\subsection{Calculating Synchrotron/Jitter Emission from Electron Trajectories in Self-consistently
Generated Magnetic Fields}
\vspace*{-0.5cm}

In order to obtain the spectrum of synchrotron (jitter) emission, we consider an ensemble of electrons selected in the region where the Weibel instability has fully grown and electrons are accelerated in the generated magnetic fields.  In order to validate our numerical method we performed simulations using a small system with ($L_{\rm x}, L_{\rm y}, L_{\rm z}) = (645\Delta, 131\Delta, 131\Delta)$ ($\Delta = 1$: grid size) and a total of $\sim 0.5$ billion particles (12 particles$/$cell$/$species for the ambient plasma) in the active grid zones \citep{nishi06}. First we performed simulations without calculating radiation up to $t = 450\omega_{\rm pe}^{-1}$ when the jet front is located at about $x/\Delta = 480$. We randomly selected 12,150 electrons in both the jet and in the ambient medium.  Recently, a similar calculations have been carried out for the radiation from electrons accelerated in laser-wakefield acceleration \citep{martins09}, in    counter-streaming jets \citep{fred10},  and from a single shock in the contact discontinuity frame \citep{sironi09j}.

Figure 3 shows (a) the current filaments generated by the Weibel instability and (b) the 
phase space of $x/\Delta - \gamma V_{\rm x}$ for jet electrons (red) and ambient electrons 
(blue) at  $t = 450\omega_{\rm pe}^{-1}$. Figure 4 shows (a) the $x$-component of current density generated by the Weibel instability  and (b) the phase space of jet electrons and ambient electrons at a slightly later time 
 $t = 525\omega_{\rm pe}^{-1}$.
\begin{figure}[h]
\begin{center}
\includegraphics[width=288.3pt, height=100pt]{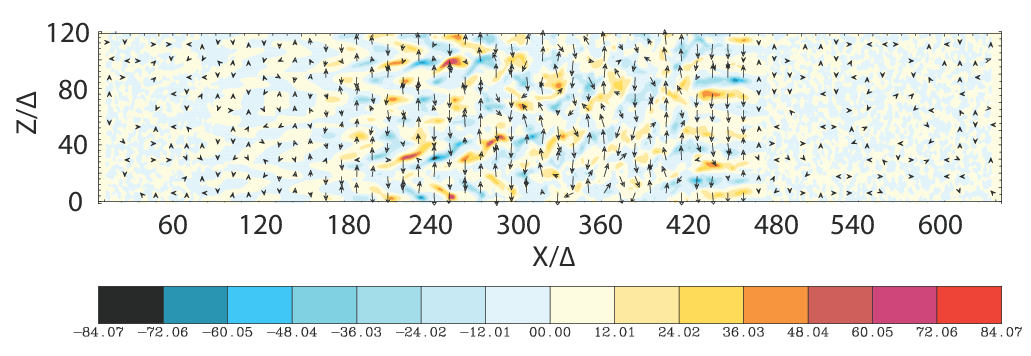} 
\includegraphics[width=177.6pt, height=100pt]{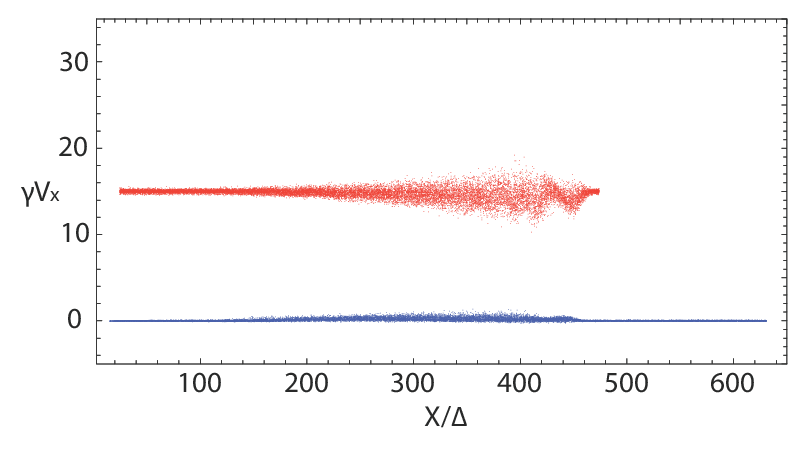}
\end{center}
\caption{\baselineskip 12pt 
Two-dimensional images in the $x-z$ plane at $y/\Delta = 65$ for $t =450 \omega_{\rm pe}^{-1}.$ 
The colors indicate the x-component of current density generated by the Weibel instability, with
the x- and z-components of magnetic field represented by arrows (a). Phase space distributions 
as a function of $x/\Delta-\gamma v_{\rm x}$ plotted for the jet (red) and ambient  (blue)
electrons at the same time.}
\end{figure}
\begin{figure}[h]
\begin{center}
\includegraphics[width=288.3pt, height=100pt]{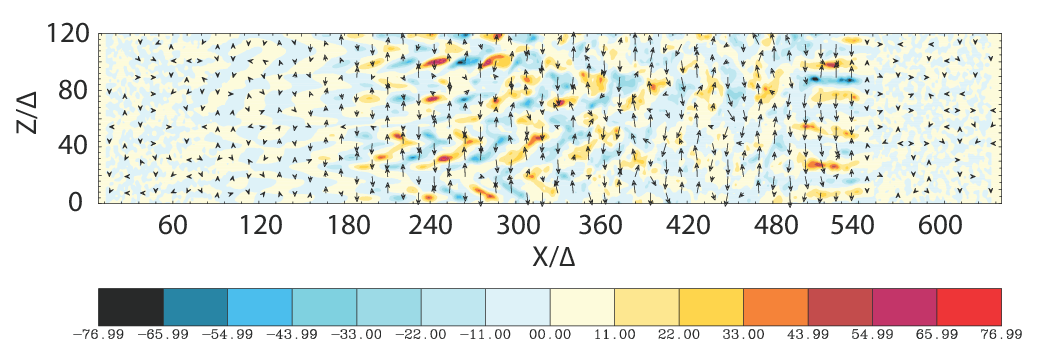} 
\includegraphics[width=177.6pt, height=100pt]{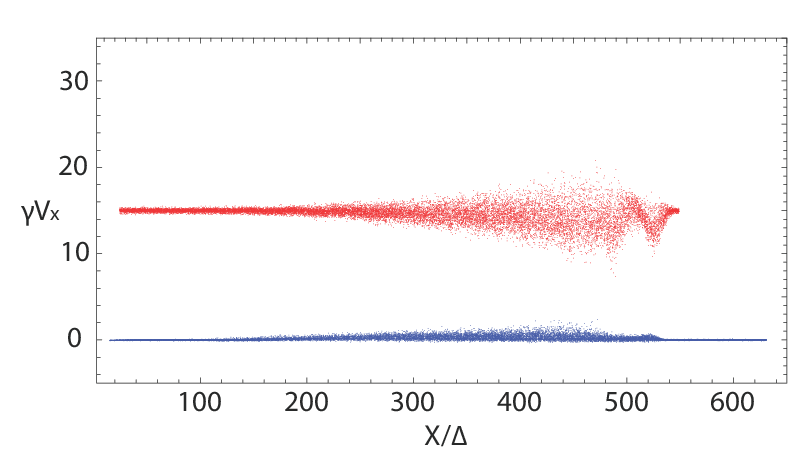}
\end{center}
\caption{\baselineskip 12pt 
Two-dimensional images in the $x-z$ plane at $y/\Delta = 65$ for $t =525 \omega_{\rm pe}^{-1}.$ 
The colors indicate
the x-component of current density generated by the Weibel instability, with
the x- and z-components of magnetic field represented by arrows (a).
Phase space distributions as a function of $x/\Delta-\gamma v_{\rm x}$ plotted for the jet (red) and 
ambient  (blue) electrons at the same time.}
\end{figure}

We calculated the emission from the jet and ambient 12,150 electrons during the sampling time $t_{\rm s} = t_{\rm 2} - t_{\rm 1} = 75\omega_{\rm pe}^{-1}$ with Nyquist frequency  $\omega_{\rm N} = 1/2\Delta t = 200\omega_{\rm pe}$ where $\Delta t = 0.005\omega_{\rm pe}^{-1} $ is the simulation time step and the frequency resolution $\Delta \omega = 1/t_{\rm s} = 0.0133\omega_{\rm pe}$. The resulting spectra shown in Figure 5 show emission from jet electrons and ambient electrons separately, and are calculated for head-on ($0^{\circ}$) and $5^{\circ}$ viewing directions. The radiation from jet electrons shows Bremsstrahlung-like spectra as the red ($0^{\circ}$) and orange ($5^{\circ}$) lines \citep{hedeT05}. The jet electron spectra are 
different from the spectra shown in Fig.\ 2c because the jet electrons are not much accelerated, the magnetic fields generated by the Weibel instability are rather weak and the electron trajectories are only slightly 
bent.  
\begin{figure}[h]
\begin{center}
\includegraphics[width=274pt, height=156pt]{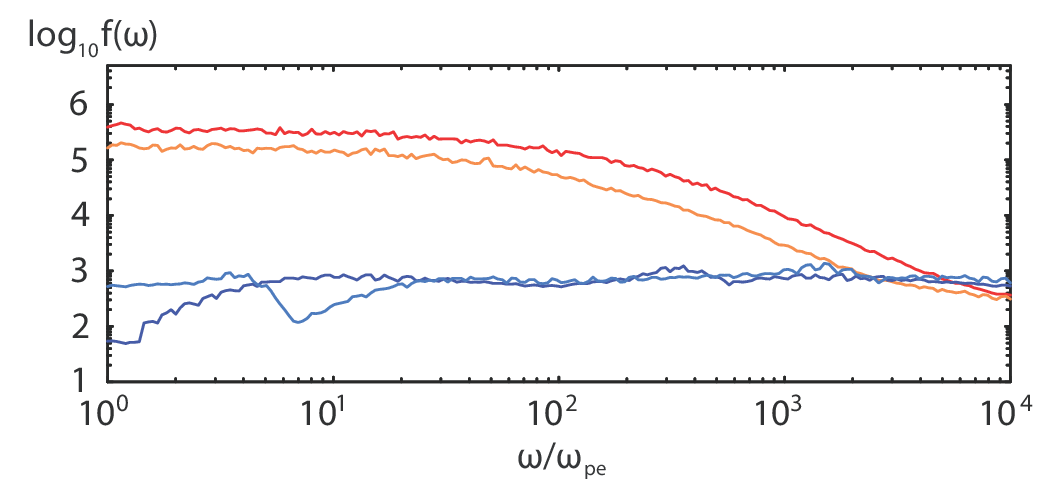} 
\end{center}
\caption{\baselineskip 12pt 
Spectra obtained from jet and ambient electrons for the two viewing angles.  Spectra for jet electrons are shown by red ($0^{\circ}$) and orange ($5^{\circ}$) lines. Spectra for ambient electrons are shown by blue ($0^{\circ}$) and light blue ($5^{\circ}$) lines.}
\end{figure}

We can compare the spectra in Fig.\ 5 with spectra obtained for two electrons as shown in Fig.\ 6.  Here we have a parallel magnetic field, $B_{\rm x} =0.37$, a jet velocity of $v_{\rm j1,2} = 0.99c$, and
two electrons with different perpendicular velocities ($v_{\perp 1} = 0.01c, v_{\perp 2} = 0.012c$). A maximum Lorentz factor of $\gamma_{\max} =\{(1 - (v_{\rm j2}^{2} +v_{\perp 2}^{2})/c^{2}\}^{-1/2}  = 7.114$ accompanies the larger perpendicular velocity. The critical angle for off-axis radiation $\theta_{\gamma} = 180^{\circ}/(\pi \gamma_{\max})$ for this case is 8.05$^{\circ}$. 
\begin{figure}[h]
\begin{center}
\includegraphics[width=110pt, height=110pt]{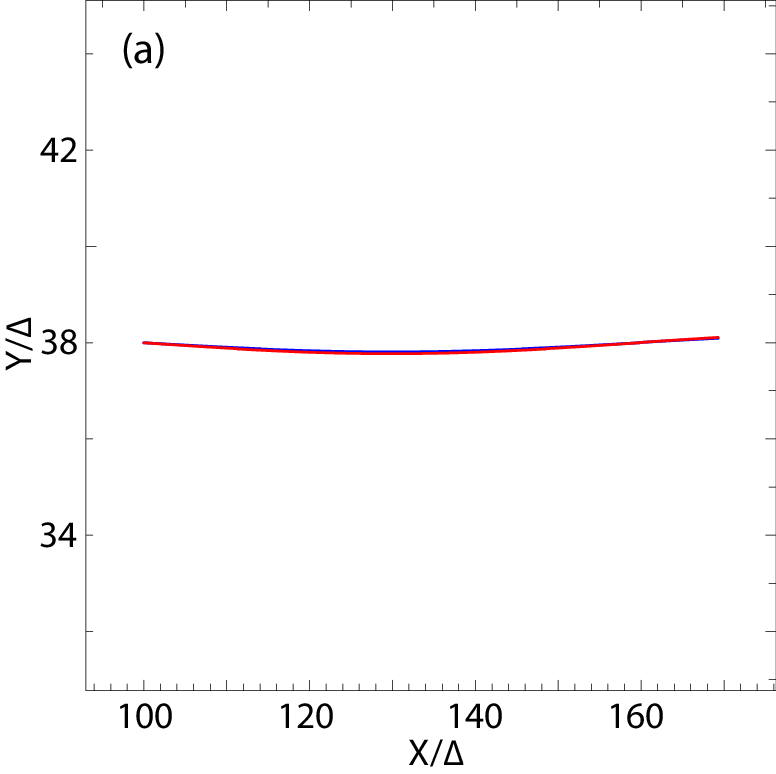}
\includegraphics[width=150pt, height=61pt]{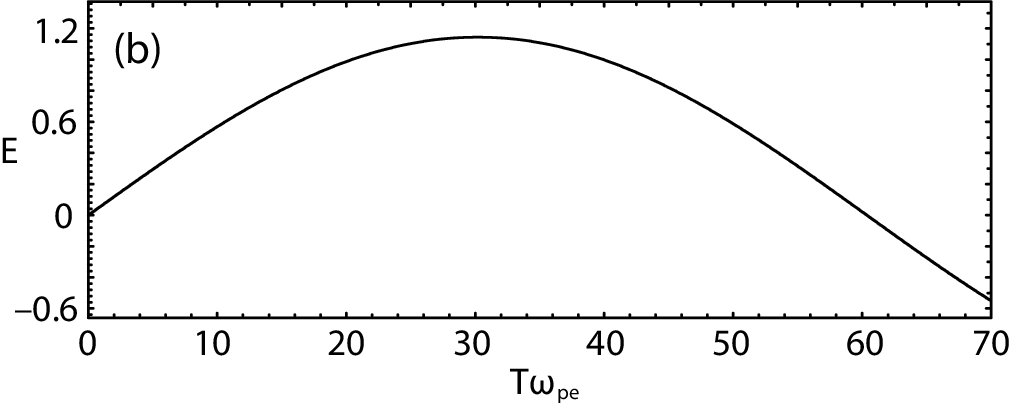}
\hspace*{0.0cm}
\includegraphics[width=120pt, height=120pt]{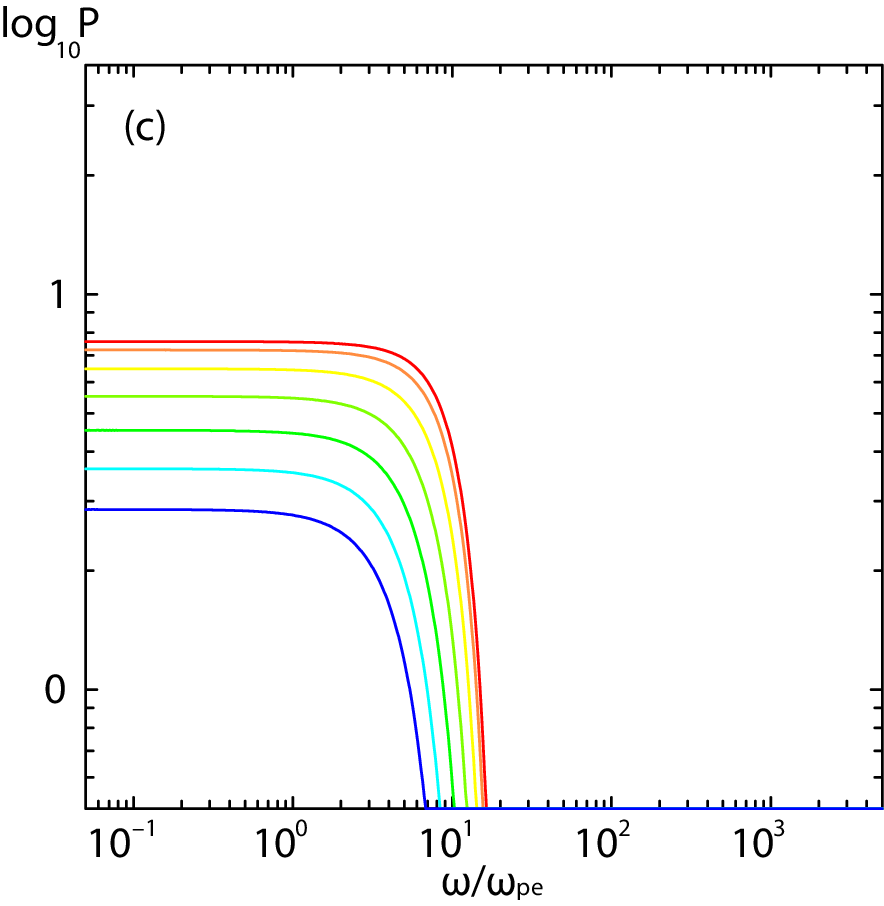}
\end{center}
\vspace*{-0.cm}
\caption{\baselineskip 12pt The case with a weak magnetic field ($B_{\rm x} = 0.37$) and small 
perpendicular velocity  ($v_{\perp 1} = 0.01c, v_{\perp 2} = 0.012c$).
The paths of two electrons moving helically along the $x-$direction in a homogenous magnetic
field shown in the $x-y$-plane (a). The two electrons radiate a time dependent electric field. 
An observer situated at great distance along the n-vector sees the retarded electric field from
the moving electrons (b). 
The observed power spectrum at different viewing angles from the two electrons (c). 
Frequency is in units of $\omega_{\rm pe}$.  It should be noted that the cyclotron 
frequency is around 20$\omega_{\rm pe}^{-1}$. }
\end{figure}
{\b A comparison between Fig.\ 5 and Fig.\ 6c indicates similarities. The lower frequencies have flat 
spectra and the higher frequencies decrease monotonically. The slope in Fig.\ 5 is less steep than 
in Fig.\ 6c. This is due to the fact that the spread of Lorentz factors of the jet electrons is larger and 
the average Lorentz factor is larger as well. Although the magnetic field is not as strong in the simulation 
spectra,  the spectra are extended to higher frequency. 
This is explained that as shown in Fig. 7.16 (left) in Hededal?s Ph. D. thesis (Hededal, 2005) the turbulent magnetic field shifts the frequency higher with shorter wave length (smaller $\mu$).
We obtained results for other simulations with  different parameters for the jet electrons and including an ambient magnetic field. In all cases the strength of the magnetic fields generated by the Weibel instability was small, and the spectra for these cases were very similar to Fig.\ 5.  The low level for the magnetic field energy in these small test case simulations with $x/\Delta < 600$ is to be expected. As indicated by Fig.\ 1b, the magnetic field energy in the region $x/\Delta < 500$ is small ($\epsilon_{\rm B} < 0.07$), therefore, as expected, the spectra should look like that from electrons propagating in a turbulent magnetic field with some high frequency enhancement.}

%Furthermore, even the magnetic field strength 
%is not so large, however the slope of the spectra seems to be consistent with Fig. 7.16 (left) with 
%the turbulent magnetic field with the red noise ($\mu = -3$) in Hededal's
%Ph. D. thesis \citep{hedeT05}. We obtained several different parameters with 
%jet electrons and ambient magnetic field. However, the strength of the magnetic 
%fields generated by the Weibel instability is small, therefore the spectra for 
%these cases are very similar to Fig. 5. As shown in Fig. 7.12 in Hededal's Ph. D. 
%thesis \citep{hedeT05}, the trajectories of jet electrons have to be chaotic to 
%produce a jitter-like spectrum  Fig. 7.22 \citep{hedeT05}.

%As shown in Fig. 1b, the magnetic field energy in the region $x/\Delta < 500$ is 
%small ($\epsilon_{\rm B} < 0.07$), therefore, as expected, the spectra look like 
%those emitted from electrons propagating in a turbulent magnetic field with red noise
%(see also Figs. 3a and 4a).

\vspace*{-0.9cm}
\section{Discussion}
\vspace*{-0.5cm}

Emission obtained with the method described above is self-consistent, and automatically accounts for magnetic field structures on the small scales responsible for jitter emission. By performing such calculations for simulations with different initial parameters, we can investigate and compare the different regimes of jitter- and synchrotron-type emission \citep{eps73,medv00,medv06}. 
% (Medvedev 2000). 
The feasibility of this approach has already been demonstrated \citep{hedeT05,hedeN05}, 
%(Hededal \& Nordlund 2005; Hededal 2005), 
and its implementation is straightforward. Thus, we should be able to address the low 
frequency GRB spectral index violation of the synchrotron spectrum line of death \citep{medv06}. 
% (Medvedev 2006a).

 Recently, synthetic radiation has been obtained  from electrons accelerated in laser-wakefield acceleration \citep{martins09}, in   counter-streaming jets \citep{fred10}, and from a single shock in the contact discontinuity frame \citep{sironi09j}.  Our simulation setup is different from these simulations in that we do not have a fixed contact discontinuity (CD) (reflected at the wall) \citep[e.g.][]{sironi09j}, or counter-streaming jets  \citep[e.g.][]{martins09,fred10}.  Instead we have been performing \citep[e.g.][]{nishi09} simulations like that shown in Fig.\ 1, where we inject relativistic jets into an ambient plasma. A double shock structure (bow and jet shocks separated by a contact discontinuity region) is formed and electrons can be accelerated due to the Weibel instability in both shocks. Since we calculate the radiation from the electrons in the observer frame, and calculated spectra can be compared directly with observations. As shown in Fig.\ 1, the strongest electron acceleration and strongest magnetic fields are generated in the jet (trailing) shock. Therefore, in this simulation this region would produce the emission that is observed.

Medvedev and Spitkovsky recently showed that electrons may cool efficiently at or near the shock 
jump and are capable of emitting a large fraction of the shock energy \citep{MS09}. Such shocks 
are well-resolved in existing PIC simulations; therefore, the microscopic structure can be studied
in detail. Since most of the emission in such shocks would originate from the vicinity of the shock, the spectral power of the emitted radiation can be directly obtained from finite-length simulations and compared with observational data. We will calculate more spectra from RPIC simulations and compare in detail with Fermi data in future work.

%%%%%%%%%%%%%%%%%%%%%%%%%%%%%%%%%%%%%%%%%%%%%%%%
%% BACKMATTER
%%%%%%%%%%%%%%%%%%%%%%%%%%%%%%%%%%%%%%%%%%%%%%%%
\vspace*{-0.9cm}
%\begin{theacknowledgments}
\section{Achknowledgments}
\vspace*{-0.5cm}
This work is supported by NSF-AST-0506719, AST-0506666, AST-0908040, AST-0908010, NASA-NNG05GK73G, NNX07AJ88G, NNX08AG83G, NNX  08AL39G, and NNX09AD16G.  JN was supported by MNiSW research projects 1 P03D 003 29 and N N203 393034, and The Foundation for Polish Science through the HOMING program, which is  supported through the EEA Financial Mechanism.Simulations were performed at the  Columbia facility at the NASA Advanced Supercomputing (NAS).  and IBM p690 (Copper) at the National
Center for Supercomputing Applications (NCSA) which is supported by the NSF. Part of this work was done while K.-I. N. was visiting the Niels Bohr Institute. Support from the Danish Natural Science Research Council is gratefully acknowledged. This report was finalized during the program ``Particle Acceleration in Astrophysical Plasmas'' at the Kavli Institute for Theoretical Physics which is supported by  the National Science Foundation under Grant No.\ PHY05-51164.

%\end{theacknowledgments}



\bibliographystyle{aipproc}   % if natbib is available

\vspace*{-0.9cm}
\begin{thebibliography}{}
\vspace*{-0.7cm}

\bibitem[Achterberg et al.(2001)]{Achterberg01}
 Achterberg,  A., Gallant, Y. A.  Kirk, J. G., \&    Guthmann, 
A. X., Particle acceleration by ultrarelativistic shocks: theory and simulations, 
MNRAS, 328, 393 - 408,  2001.

\bibitem[Bekefi(1966)]{beke66}
 Bekefi, G., {\em Radiative Processes in Plasmas}, John Wiley \& Sons, New York, 1966.

\bibitem[Chang,   Spitkovsky,  \&  Arons(2008)]{chang08}
 Chang, P.,  Spitkovsky, A., \&  Arons, J., Long-term evolution of magnetic turbulence in relativistic
 collisionless shocks: Electron-positron plasmas,
  ApJ,  674, 378 - 387, 2008.

\bibitem[Ellison \& Double(2002)]{ellison02}
 Ellison,  D. C., \&  Double, G. P., Nonlinear particle acceleration in relativistic shocks, 
Astroparticle Phys.,  18, 213 - 228,
2002.

\bibitem[Epstein(1973)]{eps73}
Epstein, R. I., Synchrotron sources. I. Extension of theory for small pitch angles, ApJ, 183, 593 - 610, 1973.

\bibitem[Frederiksen et al.(2010)]{fred10}
Frederiksen, J. T., Haugb\o lle, T., Medvedev, M. V.,  \&   Nordlund,  \AA., 
Radiation spectral synthesis of relativistic filamentation, ApJ, 722, L114 - L119, 2010.

\bibitem[Hededal(2005)]{hedeT05}
 Hededal, C.B.,  Ph.D. thesis, Gamma-Ray Bursts, Collisionless Shocks and Synthetic Spectra, 2005. (arXiv:astro-ph/0506559)

\bibitem[Hededal \& Nordlund(2005)]{hedeN05}
 Hededal, C.B.,  \&   Nordlund,  \AA.,  Gamma-ray burst synthetic spectra from collisionless shock PIC
 simulaitons,  ApJL, submitted, 2005.
(arXiv:astro-ph/0511662)

\bibitem[Hededal \& Nishikawa(2005)]{hede05}  Hededal, C. B.,  \&  Nishikawa, K.-I.,
 The influence of an Ambient
   Magnetic field on Relativistic Collisionless Plasma Shocks, ApJ, 623, L89 - L92,  2005.

\bibitem[Jackson(1999)]{jack99}
 Jackson, J. D., {\em Classical Electrodynamics}, Interscience, 1999.

\bibitem[Jaroschek \& Hoshino(2009)]{jaro09}
Jaroschek C. H. \& Hoshino,  M., Radiation-Dominated Relativistic Current Sheets, PRL, 103, 075002 1-4,
2009. 

\bibitem[Landau \& Lifshitz(1980)]{land80}
 Landau, L. D., \&  Lifshitz, E. M.,{\em The Classical Theory of Fields}, Elsevier Science \& Technology Books,
1980.

\bibitem[Lemoine \& Pelletier(2003)]{lemoi03}  Lemoine,  M., \&
 Pelletier, G., Particle Transport in Tangled Magnetic Fields and Fermi Acceleration at Relativistic Shocks, 
ApJ,  589, L73 - L76, 2003.

\bibitem[Martins et al.(2009)]{martins09}
Martins, J.L.,  Martins, S.F.,  Fonseca, R.A.   Silva, L.O.,
Radiation post-processing in PIC codes
Proc. of SPIE, 7359, 73590V-1 - 8,  2009.

\bibitem[Medvedev(1999)]{medv99}
 Medvedev,  M.~V., \&   Loeb, A., Generation of magnetic fields in the relativistic shock 
 of gamma-ray burst sources, ApJ, 526, 697 - 706, 1999.

\bibitem[Medvedev(2000)]{medv00}
 Medvedev, M. V.,  Theory of ``Jitter'' Radiation from Small-Scale Random Magnetic Fields and Prompt Emission from Gamma-Ray Burst Shocks,  
ApJ, 540, 704 - 714, 2000.

\bibitem[Medvedev(2006)]{medv06}
Medvedev, M. V.,  The theory of spectral evolution of the GRB prompt emission, ApJ, 637, 869 - 872, 2006.

\bibitem[Medvedev \& Spitkovsky(2009)]{MS09}
 Medvedev, M.V., \&     Spitkovsky, A., Radiative cooling in relativistic collisionless shocks. Can simulations and experiments probe relevant GRB physics?,   
ApJ, 700, 956 - 964,  2009. 

\bibitem[Meszaros(2002)]{mes02}
Meszaros, P., Theories of Gamma-Ray Bursts,  
ARAA,  40, 137 - 169, 2002.

\bibitem[Meszaros(2006)]{mes06}
 Meszaros, P.,  Gamma-Ray Bursts, 
Rept. Prog. Phys.,  69, 2259 - 2321,  2006.

\bibitem[Meszaros \& Rees(1997)]{mr97}
Meszaros,  P., \&   Rees, M. J., Poynting Jets from Black Holes and Cosmological Gamma-Ray Bursts,  
ApJ,  482, L29 - L32, 1997.

\bibitem[Meszaros, Rees, \& Wijer(1998)]{mrw98}
 Meszaros, P., Rees, M. J., Wijers, R. A. M. J., Viewing Angle and Environment Effects in Gamma-Ray Bursts: Sources of Afterglow Diversity, 
ApJ, 499, 301 - 308, 1998.

\bibitem[Nakar(2007)]{nakar07}
 Nakar, E., Short-hard gamma-ray bursts, 
Phys. Reports, 442, 166 - 236, 2007.

\bibitem[Niemiec, \& Ostrowski(2006)]{niemiec06}
 Niemiec, J., \&     Ostrowski,  M.,   Cosmic Ray Acceleration at Ultrarelativistic Shock Waves: Effects of
 a ``Realistic'' Magnetic Field Structure, ApJ, 641, 984 - 992, 2006.

\bibitem[Niemiec, Ostrowski, \& Pohl(2006)]{niemiecp06}
 Niemiec, J.,  Ostrowski,  M.,  Pohl, M.,  Cosmic-Ray Acceleration at Ultrarelativistic Shock Waves: Effects of
Downstream Short-Wave Turbulence, ApJ, 650, 1020 - 1027,  2006.


\bibitem[Nishikawa et al.(2003)]{nishi03}
Nishikawa, K.-I., Hardee, P., Richardson, G., Preece, R.,   Sol, H.,
\&  Fishman, G.~J.,  Particle Acceleration in Relativistic Jets due
    to Weibel Instability, ApJ, 595, 555 - 563, 2003.

\bibitem[Nishikawa et al.(2005)]{nishi05}
 Nishikawa, K.-I.,  Hardee, P., Richardson, G., Preece, R., Sol, H.,
\&  Fishman, G.~J., Particle Acceleration and Magnetic Field Generation
    in Electron-Positron Relativistic Shocks, ApJ, 623, 927 - 937, 2005.

\bibitem[Nishikawa et al.(2006)]{nishi06}
 Nishikawa, K.-I.,  Hardee, P.,  Hededal,  C.~B., \&  Fishman, G.~J., Acceleration Mechanics in 
 Relativistic Shocks by the Weibel Instability, ApJ, 642, 1267 - 1274, 2006.

\bibitem[Nishikawa et al.(2008)]{nishi08}
Nishikawa, K. -I.,  Niemiec, J.,   Sol, H.,   Medvedev, M.,  Zhang,  B.,  Nordlund, \AA.,   
Frederiksen,  J. T.,  Hardee, P.,  Mizuno, Y.,  Hartmann, D., \&  Fishman,  G. J.,
New Relativistic Particle-In-Cell Simulation Studies of Prompt and Early Afterglows from GRBs,  in Proceedings of 
  {\em The 4th Heidelberg International Symposium on High Energy Gamma-Ray Astronomy},  eds, 
 F. A. Aharonian, W. Hofmann, F. Rieger
1085, 589 - 593,  2009.
 (arXiv:0809.5067)

\bibitem[Nishikawa et al.(2009)]{nishi09}
Nishikawa,  K.-I.,   Niemiec, J.,   Hardee, P. E.,  Medvedev, M.
Sol, H.,  Mizuno, Y.,   Zhang, B.,
 Pohl, M., Oka,  M., \&   Hartmann, D. H.,  Weibel instability and associated strong fields
in a fully 3D simulation of a relativistic shock, ApJ, 689, L10 - L13, 2009.

\bibitem[Panaitescu,  \& Kumar(2001)]{paku01}  Panaitescu, A., \&    Kumar, P., Fundamental Physical Parameters of Collimated Gamma-Ray Burst Afterglows, 
ApJ, 560, L49 - L52,  2001.

\bibitem[Piran(1999)]{p99}
Piran, T., Gamma-ray bursts and the fireball model, 
Phys. Rep., 314, 575 - 667, 1999.

\bibitem[Piran(2000)]{p00}
Piran, T., Gamma-ray bursts - a puzzle being resolved, 
Phys. Rep., 333, 529 - 553,  2000.

\bibitem[Piran(2005a)]{p05a}  Piran, T., The physics of gamma-ray bursts, 
Rev. Mod. Phys., 76,
 1143 - 1210, 2005a.

\bibitem[Piran(2005b)]{p05b}  Piran, T., Magnetic Fields in Gamma-Ray Bursts: A Short Overview,
 in the proceedings of
{\em Magnetic Fields in the Universe}, Angra dos Reis, Brazil, Nov. 29-Dec
3, 2004, Ed. E. de Gouveia del Pino, AIPC, 784, 164 - 174, 2005b.

\bibitem[Preece et al.(1998)]{pre98}
 Preece, R. D.,  Briggs,  M. S.,   Mallozzi, R. S.,    Pendleton, G. N.,
   Paciesas, W. S., \&  Band, D. L., The Synchrotron Shock Model Confronts a ``Line of Death'' in the BATSE Gamma-Ray Burst Data,  
 ApJ, 506, L23 - L26, 1998.

\bibitem[Ramirez-Ruiz,   Nishikawa,  \&  Hededal(2007)]{ram07}
 Ramirez-Ruiz, E.,  Nishikawa, K.-I., \&  Hededal,  C.~B., e$^{\pm}$
Loading and the origin of the upstream magnetic field in GRB shocks,
  ApJ, 671, 1877 - 1885, 2007.

\bibitem[Reville, \& Kirk(2010)]{rev10}
Reville, B. \& Kirk, J. G., Computation of Synthetic Spectra from Simulations of Relativistic Shocks, ApJ, 724, 1283 - 1295, 2010.

\bibitem[Rybicki \& Lightman(1979)]{rybi79}
Rybicki, G. B., \&  Lightman,  A. P., {\em Radiative Processes in Astrophysics}, John Wiley \& Sons, New York,
1979.

\bibitem[Sari, Piran,  \& Narayan(1998)]{spn98}
 Sari, R.,   Piran,   T., Narayan, R., Spectra and Light Curves of Gamma-Ray Burst Afterglows, 
ApJ, 497, L17 - L20,
1998.

 \bibitem[Sironi \& Spitkovsky(2009a)]{sironi09m}
Sironi, L.,  \&  Spitkovsky,  A., Particle acceleration in relativistic magnetized collisionless paslas:
Dependence of shock acceleration on magnetic obliquity,  ApJ, 698, 1523 - 1549, 2009a.

\bibitem[Sironi \& Spitkovsky(2009b)]{sironi09j}
Sironi, L. \&   Spitkovsky, A.,
Synthetic  Spectra from PIC simulations of relativistic collisionless shocks,  ApJ, 707, L92ÐL96, 2009b.

\bibitem[Spitkovsky(2008a)]{anat08a}
 Spitkovsky, A.,  On the structure of relativistic collisionless shocks in electron-ion plasmas, 
 ApJ,  673, L39 - L42, 2008a.
 
\bibitem[Spitkovsky(2008b)]{anat08b}
 Spitkovsky,  A.,  Particle acceleration in relativistic collisionless shocks: Fermi process  
 at last, ApJ,  682, L5 - L8, 2008b.

\bibitem[Waxman(2006)]{wax06}  Waxman, E., Gamma-ray bursts and collisionless shocks,  
Plasma Phys. Control. Fusion,
 48,  B137 - B151, 2006.  (arXiv:astro-ph/0607353)	

\bibitem[Yost et al.(2003)]{yost03}
 Yost, S. A.,   Harrison, F. A.,  Sari, R.,  Frail, D. A., A Study of the Afterglows of Four Gamma-Ray Bursts: Constraining the Explosion and Fireball Model, 
ApJ, 597, 459 - 473,  2003.

\bibitem[Zhang \& Meszaros(2004)]{zm04}
 Zhang, B., \&    Meszaros, P., Gamma-Ray Bursts: progress, problems \& prospects, 
 Int. J. Mod. Phys.,  A19, 2385 - 2472,
2004.  

\bibitem[Zhang(2007)]{zhangr07}
 Zhang,  B., Gamma-Ray Bursts in the Swift Era, 
	Chin. J. Astron. Astrophys.,  7, 1 - 50, 2007.
	
\end{thebibliography}


%\end{document}

%%%%%%%%%%%%%%%%%%%%%%%%%%%%%%%%%%%%%%%%%%%
%% The following lines show an example how to produce a bibliography
%% without the help of the BibTeX program. This could be used instead
%% of the above.
%%%%%%%%%%%%%%%%%%%%%%%%%%%%%%%%%%%%%%%%%%%

\end{document}